\input{aipcheck}

\documentclass[
    ,final            
  ]
  {aipproc}

\layoutstyle{6x9}

\usepackage{amssymb,amsmath}

\begin{document}

\title{Direct Photons in Heavy-Ion Collisions}


\author{Klaus Reygers}{
  address={University of M{\"u}nster, Institut f{\"u}r Kernphysik, \\
           Wilhelm-Klemm-Stra{\ss}e 9, 48149 M{\"u}nster, Germany}
}



\begin{abstract}
  A brief overview of direct-photon measurements in ultra-relativistic
  nucleus-nucleus collisions is given. The results for Pb+Pb
  collisions at $\sqrt{s_\mathrm{NN}} = 17.3$~GeV and for Au+Au
  collisions at $\sqrt{s_\mathrm{NN}} = 200$~GeV are compared to
  estimates of the direct-photon yield from hard scattering. Both
  results leave room for a significant thermal photon component. A
  description purely based on hard scattering processes, however, is
  not ruled out so far.
\end{abstract}

\maketitle


\section{Introduction}
In ultra-relativistic heavy-ion collisions it is expected that for a
brief period of several fm/$c$ a thermalized medium is created whose
relevant degrees of freedom are quarks and gluons. It has long been
suggested that the initial temperature of this quark-gluon plasma
(QGP) can be determined via the measurement of direct photons, {\it
  i.e.}, photons not coming from late hadron decays like $\pi^0
\rightarrow \gamma \gamma$ \cite{Stankus:2005eq}. The virtue of direct
photons is that they escape the hot and dense medium unscathed. The
experimental challenge is to extract a direct-photon signal above the
large decay-photon background and to identify other sources of direct
photons which are not of thermal origin.

\begin{figure}[t]
  \includegraphics[height=.18\textheight]{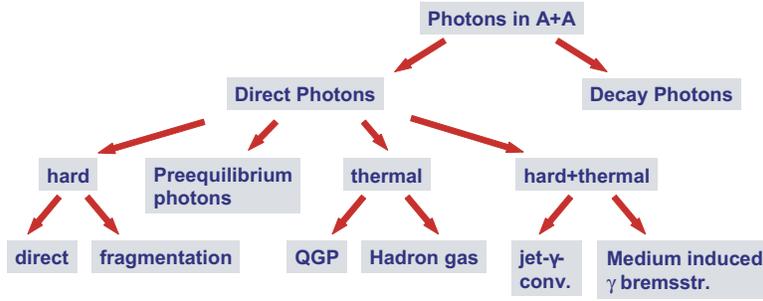}
\caption{Known and presumed photon sources in nucleus-nucleus collisions.}
\label{fig:photon_sources}
\end{figure}

A brief summary of known and presumed photon sources in
nucleus-nucleus collisions is given in Fig.~\ref{fig:photon_sources}
\cite{Turbide:2005fk}.  Photons from hard scattering of quarks and
gluons, analogous to the production mechanisms in p+p-collisions,
dominate the direct-photon spectrum at high transverse momenta
($p_\mathrm{T}$). The main motivation for the measurement of
high-$p_\mathrm{T}$ photons in heavy-ion collisions is to test
perturbative QCD models and to measure the rate of initial hard
scatterings.

The QGP expands and cools and at a temperature of $T_\mathrm{c}
\approx 190$~MeV a phase transition to a hadron gas takes place
\cite{Cheng:2006qk}.  During the entire evolution of the QGP and the
hadron gas thermal direct photon are produced. The shape of their
$p_\mathrm{T}$ spectra reflects the temperature of the medium.
Thermal photon are expected to contribute to the direct photon
spectrum significantly at low $p_\mathrm{T}$ ($\lesssim 3$~GeV/$c$).
For model comparisons and the extraction of the initial temperature
model calculations need to convolve photon rates for the QGP and the
hadron gas with realistic scenarios of the space-time evolution of the
fireball. Initial temperatures $T_\mathrm{i} > T_\mathrm{c}$ would
provide evidence for the creation of a QGP.

Direct photons might furthermore be produced in interactions of quarks
or gluons from early hard scattering processes with soft quarks and
gluons from the QGP. One suggested mechanism is jet-photon conversion
in processes like $q_\mathrm{hard} + g_\mathrm{QGP} \rightarrow \gamma
+ q$ and $q_\mathrm{hard} + \bar{q}_\mathrm{QGP} \rightarrow \gamma +
g$ in which the photon obtains a large fraction of the momentum of
$q_\mathrm{hard}$ \cite{Fries:2002kt}.  In Au+Au collisions at
$\sqrt{s_\mathrm{NN}} = 200$~GeV jet-photon conversion might be a
significant direct-photon source for $p_\mathrm{T} \lesssim 6$~GeV/$c$.
Direct photons might furthermore be produced due to multiple
scattering of quarks in the medium. These interesting ideas, however,
still require a experimental verification.

\section{Measurements: WA98 and PHENIX}
\begin{figure}[th]
\includegraphics[height=.31\textheight]{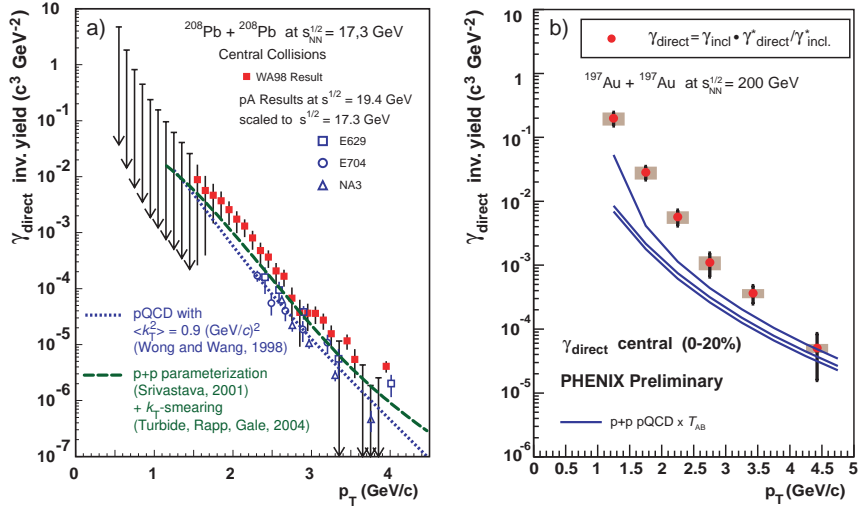}
\caption{Direct-photon spectra measured in Pb+Pb collisions at the 
  CERN SPS (WA98 experiment) and in Au+Au collisions at RHIC (PHENIX
  experiment). Both spectra are compared to estimates of the
  contribution of hard scattering processes in order determine whether
  an additional thermal photon contribution is needed.}
\label{fig:wa98_phenix}
\end{figure}

Direct photons were measured by the fixed-target experiment WA98 at
the CERN SPS in central Pb+Pb collisions at $\sqrt{s_\mathrm{NN}} =
17.3$~GeV \cite{Aggarwal:2000th} and by the PHENIX experiment at the
Relativistic Heavy-Ion Collider (RHIC) in Au+Au collisions at
$\sqrt{s_\mathrm{NN}} = 200$~GeV \cite{Adler:2005ig, Bathe:2005nz}
(see Fig.  \ref{fig:wa98_phenix}). One of the basic questions in both
cases is whether thermal photons or photons from jet-plasma
interactions are needed on top of the hard direct-photon component in
order to explain the data.

\begin{figure}[t]
\includegraphics[height=.28\textheight]{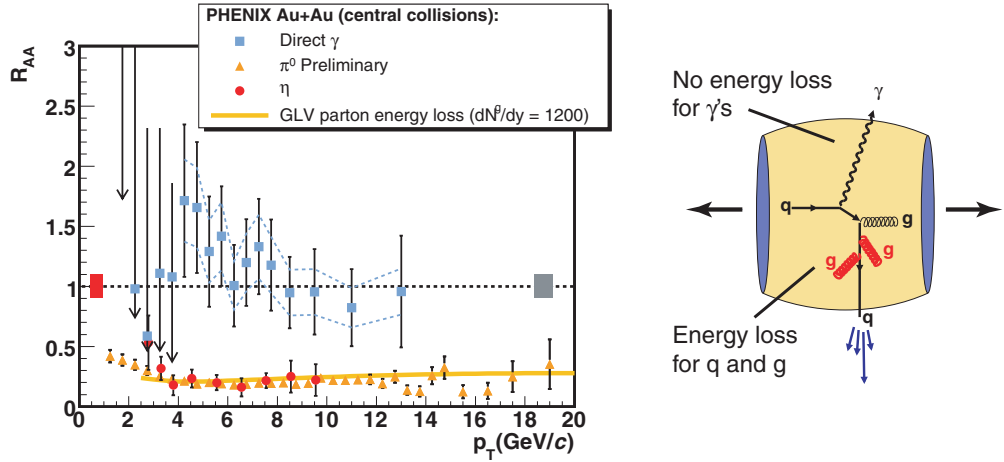}
\caption{Nuclear modification factor $R_\mathrm{AA}$ for 
  direct-photons, neutral pions, and $\eta$ mesons in central Au+Au
  collisions at $\sqrt{s_\mathrm{NN}} = 200$~GeV. Pions and
  $\eta$-mesons are suppressed whereas direct photons are not. The
  cartoon illustrates the most popular explanation: energetic quarks
  and gluons which fragment into hadrons suffer energy loss in the
  medium, direct photons don't.}
\label{fig:raa}
\end{figure}

In both experiments the direct-photon spectra are determined by a
statistical subtraction of the calculated yield of photons from hadron
decays from the total photon yield. The WA98 measurement was made with
a highly segmented lead-glass calorimeter.  PHENIX measured
high-$p_\mathrm{T}$ direct photons ($p_\mathrm{T} \gtrsim 4$~GeV/$c$)
in a similar way with its electromagnetic calorimeters (see
Fig.~\ref{fig:raa}). The preliminary low-$p_\mathrm{T}$ direct-photon
spectrum shown in Fig.~\ref{fig:wa98_phenix}b was obtained by
measuring virtual photons via their decay into of $e^+e^-$ pairs with
the aid of a Ring Imaging Cherenkov Detector.  The spectrum of real
direct photons can then be obtained under the assumption that the
fraction $\gamma_\mathrm{direct}/\gamma_\mathrm{all}$ of real direct
photons is identical to the fraction
$\gamma^*_\mathrm{direct}/\gamma^*_\mathrm{all}$ of virtual direct
photons with small mass ($\le 30$~MeV) \cite{Bathe:2005nz}.

The spectrum of direct photons in central Pb+Pb collisions at
$\sqrt{s_\mathrm{NN}} = 17.3$~GeV in Fig.~\ref{fig:wa98_phenix}a is
compared to p+p and p+A direct-photon data measured at slightly higher
$\sqrt{s_\mathrm{NN}}$. These data sets have been scaled to
$\sqrt{s_\mathrm{NN}} = 17.3$~GeV and furthermore scaled by the
respective number of nucleon-nucleon collisions in p+A and central
Pb+Pb \cite{Aggarwal:2000th}. The underlying assumption in both cases
is that direct photons in p+p and p+A are produced in hard scattering
processes.  This comparison shows that for $p_\mathrm{T} \gtrsim
2.5$~GeV/$c$ the direct-photon yield in central Pb+Pb collisions is
consistent with the expected yield from hard scattering. Another
possibility to pin down the hard scattering contribution is a
comparison to a perturbative QCD (pQCD) calculation and to a
parameterization of p+p direct-photon data.
Fig.~\ref{fig:wa98_phenix}a indicates that a thermal photon signal
might be present below $p_\mathrm{T} \approx 2.5$~GeV/$c$. However, a
solid estimate of the hard scattering contribution at CERN SPS
energies remains difficult.  Firm conclusions can only be drawn if,
{\it e.g.}, a better understanding of the modification of the hard
scattering yield in Pb+Pb due to multiple soft scattering of the
incoming partons prior to the hard process ("Cronin" or "nuclear
$k_\mathrm{T}$" effect) can be achieved.

The PHENIX low-$p_\mathrm{T}$ direct-photon spectrum in
Fig.~\ref{fig:wa98_phenix}b is compared to a next-to-leading-order p+p
pQCD calculation scaled by the number of nucleon-nucleon collisions.
The three different pQCD curves correspond to different scales used in
the calculation and reflect theoretical uncertainties. An advantage at
RHIC energies is that the modification of the hard scattering yield
due to the Cronin effect is expected to be small \cite{Adler:2006wg}.
The difference between the data and the hard scattering yield as
estimated by the pQCD calculation hints at the presence of significant
thermal photon signal. This will be confirmed or disproved with
forthcoming low-$p_\mathrm{T}$ direct-photon measurements in p+p and
d+Au at the same energy.

Despite these difficulties several attempts have been made to describe
the WA98 and PHENIX direct-photon spectra with a combination of a hard
and a thermal component and to extract the initial temperature of the
thermalized fireball. Both measurements are consistent with a QGP
scenario. Initial temperatures for central Pb+Pb collisions at
$\sqrt{s_\mathrm{NN}} = 17.3$~GeV roughly range from $200 \lesssim
T_\mathrm{i} \lesssim 370$~MeV. For central Au+Au collisions at
$\sqrt{s_\mathrm{NN}} = 200$~GeV the extracted initial temperatures
tend to be higher and cover the range $370 \lesssim T_\mathrm{i}
\lesssim 570$~MeV \cite{Reygers:2006kh}.

The high-$p_\mathrm{T}$ direct-photon measurement in central Au+Au
collisions at $\sqrt{s_\mathrm{NN}} = 200$~GeV is presented in
Fig.~\ref{fig:raa} in terms of the nuclear modification factor
\begin{equation}
R_\mathrm{AA}(p_\mathrm{T}) = 
  \frac{\mathrm{d}N/\mathrm{d}p_\mathrm{T}|_\mathrm{A+A}}
       {\langle T_\mathrm{AA}\rangle 
        \times \mathrm{d}\sigma/\mathrm{d}p_\mathrm{T}|_\mathrm{p+p}} \; .
\end{equation} 
The nuclear overlap function $T_\mathrm{AA}$ is related to the number
of inelastic nucleon-nucleon collisions according to $\langle
T_\mathrm{AA} \rangle = \langle N_\mathrm{coll} \rangle /
\sigma_\mathrm{inel}^\mathrm{NN}$.  Fig.~\ref{fig:raa} shows that
unlike pions and $\eta$-mesons high-$p_\mathrm{T}$ direct photons are
not suppressed, {\it i.e.}, they follow $N_\mathrm{coll}$ scaling as
expected for hard processes.  This is in line with jet-quenching
models which attribute the hadron suppression to energy loss of
highly-energetic quarks and gluons from initial hard scattering
processes in the QGP.

\section{Conclusions}
Direct-photon measurements in nucleus-nucleus collisions from WA98
(Pb+Pb at $\sqrt{s_\mathrm{NN}} = 17.3$~GeV) and PHENIX (Au+Au at
$\sqrt{s_\mathrm{NN}} = 200$~GeV) have been discussed.  Both
measurements are consistent with a thermal photon signal and initial
temperatures $T_\mathrm{i} > T_\mathrm{c}$.  A description purely
based on hard scattering processes, however, is not ruled out so far.
The observation that hadrons at high-$p_\mathrm{T}$ are suppressed
whereas direct photons are not supports jet-quenching models.








\bibliography{sample}

\begin{thebibliography}{9}

\bibitem{Stankus:2005eq}
  P.~Stankus,
  Ann.\ Rev.\ Nucl.\ Part.\ Sci.\  {\bf 55} (2005) 517.

\bibitem{Adler:2005ig}
  S.~S.~Adler {\it et al.}  [PHENIX Collaboration],
  Phys.\ Rev.\ Lett.\  {\bf 94}, 232301 (2005)

\bibitem{Bathe:2005nz}
  S.~Bathe  [PHENIX Collaboration],
  Nucl.\ Phys.\ A {\bf 774} (2006) 731

\bibitem{Aggarwal:2000th}
  M.~M.~Aggarwal {\it et al.}  [WA98 Collaboration],
  Phys.\ Rev.\ Lett.\  {\bf 85} (2000) 3595

\bibitem{Cheng:2006qk}
  M.~Cheng {\it et al.},
  Phys.\ Rev.\ D {\bf 74} (2006) 054507

\bibitem{Turbide:2005fk}
  S.~Turbide, C.~Gale, S.~Jeon and G.~D.~Moore,
  Phys.\ Rev.\ C {\bf 72} (2005) 014906

\bibitem{Fries:2002kt}
  R.~J.~Fries, B.~Muller and D.~K.~Srivastava,
  Phys.\ Rev.\ Lett.\  {\bf 90} (2003) 132301

\bibitem{Reygers:2006kh}
  K.~Reygers  [PHENIX Collaboration],
  arXiv:nucl-ex/0608043.

\bibitem{Adler:2006wg}
  S.~S.~Adler  [PHENIX Collaboration],
  arXiv:nucl-ex/0610036.

\end{thebibliography}



\end{document}